# Two-Dimensional Spintronic Circuit Architectures on Large Scale Graphene

Dmitrii Khokhriakov[1], Bogdan Karpiak[1], Anamul Md. Hoque[1], Saroj P. Dash[1*]

[1]Department of Microtechnology and Nanoscience, Chalmers University of Technology, SE-41296, Göteborg, Sweden

**Abstract**

Solid state electronics based on utilizing the electron spin degree of freedom for storing and processing information can pave the way for next-generation spin-based computing. However, the realization of spin communication between multiple devices in complex spin circuit geometries, essential for practical applications, is still lacking. Here, we demonstrate the spin current propagation in two-dimensional (2D) circuit architectures consisting of multiple devices and configurations using a large area CVD graphene on $SiO_2$/Si substrate at room temperature. Taking advantage of the significant spin transport distance reaching 34 μm in commercially available wafer-scale graphene grown on Cu foil, we demonstrate that the spin current can be effectively communicated between the magnetic memory elements in graphene channels within 2D circuits of Y-junction and hexa-arm architectures. We further show that by designing graphene channels and ferromagnetic elements at different geometrical angles, the symmetric and antisymmetric components of the Hanle spin precession signal can be remarkably controlled. These findings lay the foundation for the design of complex 2D spintronic circuits, which can be integrated into efficient electronics based on the transport of pure spin currents.

**Keywords:** graphene, spin circuit, spin device architecture, spin transport, Hanle spin precession.

**Corresponding authors:** Dmitrii Khokhriakov, Email: dmikho@chalmers.se;
Saroj P. Dash, Email: saroj.dash@chalmers.se

**Main**

The utilization of spin degree of freedom as a state variable for information processing and storage is expected to allow faster device operation and lower power consumption, promising for applications in spintronics, artificial intelligence, and quantum technology[1,2]. For the realization of hybrid spin-based logic and memory technologies such as spin transistors and all-spin-logic devices, one needs to develop spin circuit designs combining different spintronic concepts such as efficient spin injection, detection and long-distance spin transport in nonmagnetic channels with magnetization dynamics of nanomagnets[3–8]. Such spin circuits should be composed of multiple nanomagnetic elements interconnected with spin-coherent channel materials. In the last decade, the search for a spin interconnect material exhibiting long spin lifetime and spin diffusion length at room temperature is intensified with the investigation of spin transport in metals[9], semiconductors[10–13], topological materials[14] and graphene[15–19].

The concept of using graphene as interconnect for spin-polarized electrons comes naturally because of the outstanding electrical, mechanical and thermal properties, while its two-dimensional (2D) nature allows for large current densities. Most importantly, graphene is an excellent candidate for long-distance spin interconnect due to the low spin-orbit coupling and lack of hyperfine interaction in carbon, and its high electron mobility[15–19]. Spin transport properties with more than 10 nanosecond spin lifetime are demonstrated in hexagonal boron nitride (h-BN) encapsulated single crystals of exfoliated graphene samples at room temperature[20,21]. However, the non-scalability of producing graphene by exfoliation restricts the utilization of its excellent spin transport properties in batch-fabricated devices. Recently, using chemical vapor deposited (CVD) graphene, long-distance spin transport up to 16 μm with reasonably good spin transport parameters was demonstrated[22–25]. Despite these developments, so far, the spin transport studies in graphene are limited to a simple straight graphene stripe, and demonstration of spin circuits in complicated device geometries as required for spintronic memory and logic applications remained challenging.

Here, we demonstrate spin circuit architectures with large area graphene channels efficiently carrying and communicating the spin information between nanomagnets arranged in different complex geometries consisting of multiple devices. We take advantage of extraordinary long-distance spin transport observed in commercially available wafer-scale CVD graphene samples with transport lengths exceeding 34 µm at room temperature. The advantage of making spin circuits on graphene produced by a CVD method on Cu foil is the possibility to produce high-quality spintronic devices at low cost. Utilizing our optimized nanofabrication and patterning techniques compatible with industrial manufacturing processes, we carved complicated graphene Y-junction and hexa-arm spin circuit architectures. This experimental demonstration of spin communication in such complicated graphene devices is a milestone towards large-scale integration and development of spin-logic and memory technologies.

**Large scale CVD graphene spin circuits**

The schematic of a spin circuit composed of nanomagnetic elements interconnected with graphene channels is shown in Fig. 1a. The functionality of the circuit includes spintronic phenomena such as spin injection, transport, detection, precession, and magnetization switching processes. The CVD graphene spin circuit is prepared from a chip picked from a 4-inch wafer completely covered with graphene grown on Cu foil (from Grolltex Inc.) and subsequently transferred on a $SiO_2$/Si substrate. Figure 1b shows the pictures of the graphene-covered wafer, the chip with nanofabricated spin circuits and a representative hexa-arm device. The patterning of graphene circuit and preparation of ferromagnetic and non-magnetic contacts were done by multi-step electron beam lithography process. The final circuit consists of graphene channels in different geometries with ferromagnetic Co/$TiO_2$ contacts as the injector and detector electrodes for spin-polarized current, Ti/Au electrodes as reference contacts, while the $n^{++}$Si/$SiO_2$ is used as a global back gate (see Methods for details of the nanofabrication process). In the spin circuits, the graphene channels with lengths up to 26.6 µm and the ferromagnetic contacts having different widths between 200–400 nm were prepared. The ferromagnetic tunnel contact resistances ($R_c$) measured in the three-terminal configuration were around 3–30 kΩ, while the graphene channel resistance was measured to be $R_\square$ = 1–5 kΩ/$\square$ with field-effect mobility µ = 1000–2000 $cm^2V^{-1}s^{-1}$ at room temperature (see Supplementary Figure S1a). The quality of the graphene channel was verified by the Raman spectroscopy as shown in Supplementary Figure S1b. In five studied devices, 34 out of 36 (94.4%) ferromagnetic electrodes and 19 out of 31 (61.3%) non-magnetic electrodes were found functioning. The extracted spin polarization of ferromagnetic contacts is in the range between 3-10 %, with an average value $\bar{P} = 7 \pm 2.5\%$ (see Supplementary Notes 1 and 2).

**Spin transport in CVD graphene Y-junction circuit**

First, spin transport properties of the CVD graphene were investigated in a Y-junction circuit with channel length up to 26.6 µm at room temperature. Figures 2a,b show the optical microscopy picture and schematics of the device with the measurement geometry used to determine the spin transport parameters of the large area CVD graphene. The spin accumulation is created by passing a constant DC current (*I*) in the injector ferromagnetic contact, while the nonlocal (NL) pure spin transport voltage signal ($V_{NL}$) is detected by another ferromagnetic contact placed at a length *L* away from the injector on the graphene channel. This nonlocal spin injection and detection in lateral graphene spintronic devices is an effective method to generate a pure spin current for potential spintronic applications. We perform two kinds of spin-sensitive NL measurements in our CVD graphene spin circuits with different channel lengths and geometries: spin valve and Hanle spin precession measurements. The spin valve data give an estimate of the magnitude of induced spin polarization, while Hanle spin precession curves allow to extract parameters such as spin lifetime $\tau_s$, spin diffusion coefficient $D_s$ and spin diffusion length $\lambda_s$ in graphene.

Figures 2c and 2d show the measured spin valve and Hanle spin precession data in the nonlocal (NL) configuration in channels of different lengths (6.6, 20, 26.6 µm) and geometries in the graphene Y-junction. In this circuit, the ferromagnetic injector and detector contacts are placed parallel to each other. To perform the spin valve measurement, we sweep the in-plane magnetic field ($B_\parallel$) along the easy axis of the ferromagnetic contacts while recording the nonlocal resistance ($R_{NL} = \frac{V_{NL}}{I}$). The sharp changes in $R_{NL}$ are measured when the magnetization of the injector or detector switches to either parallel or antiparallel configuration. The in-plane external magnetic field switches the magnetization of the ferromagnetic electrodes at their respective coercive fields, which are defined by their widths. Figure 2c shows the change in $\Delta R_{NL}$ from ~0.4 Ω for *L* = 6.6 µm to 0.8 mΩ for *L* = 26.6 µm graphene channel lengths at room temperature. This data can be used to evaluate the spin diffusion length of graphene assuming equal polarization of the contacts and using the exponential decay of spin signal with distance, $R_{NL} \propto e^{-L/\lambda_s}$. Figure 2e shows the obtained fit, from which we extract $\lambda_s = 3.4$ µm. However, we note that the 26.6 µm (20 µm) channel has two (one) ferromagnetic contacts on top of the channel between injector and detector, which can reduce the spin signal due to spin absorption[26]. Moreover, these channels in the Y-junction have an additional graphene channel branching off between the injector and detector contacts, which is expected to further reduce the observed spin signal magnitude compared to the channel without branches (see Supplementary Note 3). This indicates that the obtained spin parameters are a lower bound estimation, and optimizations in the device design and fabrication process can yield even longer spin transport distances.

While the spin diffusion length is an intrinsic material property that characterizes its spin transport capacity, for practical applications, one can also be interested in the actual size of a device footprint over which the spin signal can be communicated and detected. This maximal spin transport distance would depend not only on the intrinsic material properties but also on the extrinsic ones that depend on the fabrication technology, including the amount of injected spins and the noise floor of the measurements. By extrapolating the fit in Fig. 2e to the noise level of the Y-junction device, we estimate its maximum detectable spin transport distance to be 34.1 μm. This result proves the possibility to maintain the spin polarization in branched graphene channels over large distances, which is crucial for the design of logic circuits.

Hanle spin precession measurements, in contrast to the spin valve, can be used to extract spin diffusion length and other parameters of the system from a single measurement[17]. They are performed by measuring the NL resistance $R_{NL}$ while applying an out-of-plane magnetic field (B⊥), which induces spin precession with the Larmor frequency resulting in the dephasing of the spin polarization, as shown in Figure 2d. The data are fitted with eq.1,

$$\Delta R_{NL} \propto \int_0^\infty \frac{1}{\sqrt{4\pi D_s t}} e^{-\frac{L^2}{4D_s t}} \cos(\omega_L t) e^{-\left(\frac{t}{\tau_s}\right)} dt, \quad (1)$$

where $\omega_L = \frac{g\mu_B}{\hbar} B_\perp$ is the Larmor spin precession frequency, $\mu_B$ is Bohr magneton, $L$ is the channel length, $D_s$ the spin diffusion coefficient and $\tau_s$ spin lifetime[27]. We extract the spin lifetime $\tau_s = 405 \pm 37$ ps, spin diffusion constant $D_s = 0.032 \pm 0.004$ m²s⁻¹ and the spin diffusion length $\lambda_s = 3.6 \pm 0.3$ μm. The obtained $\tau_s$ and $\lambda_s$ for different graphene channel lengths are shown in Fig. 2f,g. The spin diffusion length estimated from the Hanle measurements matches well with the $\lambda_s = 3.4$ μm extracted from the channel length dependence of the spin valve magnitude. Similar spin parameters for different channel lengths indicate that contact-induced spin relaxation is not a major source of spin scattering in our devices (see Supplementary Note 4). Both the spin valve and Hanle measurements establish that the spin current in CVD graphene can propagate to a channel length of 26.6 μm at room temperature and we estimate that spin signal should be detectable in channels even longer than 34 μm. This spin transport distance is the highest achieved so far at room temperature in CVD graphene grown on Cu foil and transferred on a 4-inch $Si/SiO_2$ wafer, which opens the possibility to produce high-quality spintronic devices at low cost. To be noted, after the preparation of this manuscript, we came across a recent publication where a large spin transport distance is also reported in CVD graphene grown on a Pt foil[28]. However, Cu foil is much cheaper than Pt, which makes the conventional Cu-based graphene growth process preferable for large-scale device production and further applications. For the Cu-based CVD graphene, our result shows a long spin transport distance up to 34 μm, which is significantly larger than the previously achieved distance of 16 μm[24]. These developments allow us to design and realize spin communication in complicated device architectures on CVD graphene. The Y-junction graphene spin circuit can possibly be applied in a spin summation device and in a spin demultiplexing technology controlled by drift current[8].

Further improvement of spin transport in graphene can be achieved by incorporating 2D materials like h-BN as an encapsulation[21,29–31] and a tunnel barrier[32–35]. Compared to metal oxides that are prone to pinholes, atomically-flat h-BN can provide better interface quality and uniform contact resistance, which can give smaller spread and larger spin polarization values reaching 65% in few-layers of CVD h-BN[36]. Additionally, the spin polarization in graphene can be controlled by integrating transition metal dichalcogenides (TMDCs)[37–39] and topological insulators (TIs)[40]; and non-magnetic spin injectors can be realized via charge-to-spin conversion in heterostructures with TMDCs[41,42], Weyl semimetals[43–45], and TIs[46]. Although scalable spintronic technology for the fabrication of such heterostructures is not realized at this stage, the recent advances towards wafer-scale growth and transfer of h-BN[47] and TMDCs[48] indicate the rapid development of the related techniques.

**Spin circuit with a graphene hexa-arm architecture**

To exploit the excellent spin transport properties of the CVD graphene, we further investigated spin transport and precession in a more complex graphene spin circuit with a hexa-arm geometry, as shown in Figures 3a, b. Figure 3c shows the NL spin valve measurements in a channel having a 60° angle between the graphene branches, where the ferromagnetic injector and detector contacts are placed perpendicularly to each corresponding branch. It should be noted that the measured spin signal amplitude is reduced due to the expected isotropic diffusion of spin current to the other four branches in the spin circuit (see Supplementary Note 3). Moreover, due to the non-collinearity between the injector and detector contact magnetizations, the measured spin valve signal amplitude shows only the projected component of the spin polarization in the channel on the detector magnetization. The opposite sign of the spin valve in comparison to the measurements shown in Fig. 2c indicates that one of the utilized contacts in the hexa-arm device has negative polarization.

To account for the total spin density in graphene, we employ Hanle spin precession experiments as shown in Figure 3d. Since the injected spins are rotated at an angle of 60° with respect to the detector, an asymmetric Hanle is observed. The Hanle spin precession signal is measured for both "parallel" and "antiparallel" alignment of the ferromagnetic injector and detector electrodes, which is defined by the positive or negative projection of the injector magnetization on the detector. We deconvolute the measured data $R_{NL}(B)$ into the symmetric and antisymmetric components using eq. 2:

$$R_{NL}^{\parallel}(B) = \frac{R_{NL}(B) + R_{NL}(-B)}{2}; \quad R_{NL}^{\perp}(B) = \frac{R_{NL}(B) - R_{NL}(-B)}{2}. \qquad (2)$$

Figures 3e-f show the obtained $R_{NL}^{\parallel}$ and $R_{NL}^{\perp}$, which correspond to the parallel ($\parallel$) and perpendicular ($\perp$) components of the injected spin polarization projected on the detector. The curves are simultaneously fit with equation (1) containing $\cos(\omega_L t)$ for the parallel component and $\sin(\omega_L t)$ for the perpendicular component. The extracted amplitudes of the components allow us to calculate the relative angle between the injector and detector contacts as $\theta' = \tan^{-1}\left(\Delta R_{NL}^{\perp} / \Delta R_{NL}^{\parallel}\right)$.

For the angular convention shown in Figure 4c, the signs of $\Delta R_{NL}^{\perp}$ and $\Delta R_{NL}^{\parallel}$ can be taken into consideration to obtain the absolute angle in degrees via $\theta = \theta' + 90° \cdot sign(\Delta R_{NL}^{\perp}) \cdot [1 - sign(\Delta R_{NL}^{\parallel})]$. The sign of the parallel component is taken as positive when the projection of injector magnetization on detector points in the same direction as the magnetization of the detector. For the perpendicular component, the sign is assumed positive when the projection of the injector magnetization on the orthogonal direction to the detector magnetization is pointing 90° counterclockwise to the detector magnetization. The extracted angles of -123 ± 3° for "parallel" and 60 ± 4° for "antiparallel" configurations correspond well to the geometry of the device, proving the validity of the method. Due to the negative polarization of one of the used contacts, the calculated angles correspond to the orientation of the injected (detected) spins rather than the injector (detector) contact magnetization, since the magnetization can be either parallel or antiparallel to the spin direction depending on the sign of contact polarization. Note that the extracted amplitude of the symmetric Hanle component ($|\Delta R_{NL}^{\parallel}|$ = 1.9–2.6 mΩ) is in agreement with the spin valve amplitude (Fig. 3c), showing that spin valve is only sensitive to the component of the spin polarization parallel to the detector magnetization[49].

Having observed and analyzed the asymmetric Hanle spin precession within one pair of injector/detector contacts, we performed experiments in a distributed spin-circuit geometry where the spin injected in one contact is measured with detectors in different graphene branches simultaneously. Figure 4a shows the schematic of different channel geometries used for Hanle spin precession measurements with graphene branches being oriented at angles of 60°, 120° and 0° to each other, respectively. Figure 4b shows the measured Hanle curves, having a conventional symmetric shape for the straight channel and asymmetric shapes for electrodes oriented at 60° and 120°. The deconvoluted symmetric and antisymmetric Hanle components for each geometry are shown in their corresponding panels. Using the contact 1 for the spin injection and contacts 2 and 3 simultaneously for the spin detection, we measure Hanle signals of different shapes and magnitudes. The difference in signal amplitude stems from the varying spin polarization properties of respective detector electrodes, and the inverted shape of Hanle curves indicates the opposite direction of magnetization between the two detectors, as also observed in the calculated angles of 123 ± 3° and -61 ± 6° respectively. Indeed, when performing the spin precession experiments between contacts 2 and 3 in a straight channel, we observe a conventional Hanle curve for antiparallel orientation of electrodes. After deconvolution, the antisymmetric component is evidently zero in the straight channel, confirming the antiparallel state with $\theta = 180 ± 4°$. All presented results are summarized in Figure 4d showing the amplitude of symmetric and antisymmetric Hanle components as a function of the angle between detector and injector contact magnetizations present in different branches of the hexa-arm graphene spin circuit. The solid lines are the theoretical curves whereas the points correspond to the experimental data normalized according to eq 3.

$$\Delta \overline{R}_{NL}^{x} = \frac{\Delta R_{NL}^{x}}{\sqrt{\left(\Delta R_{NL}^{\parallel}\right)^2 + \left(\Delta R_{NL}^{\perp}\right)^2}} \quad (3)$$

The excellent correspondence between the geometrical device contact angles and values calculated from the measurements demonstrates the validity of the model with possible applications in complex non-collinear spin circuit designs. In future, the studied hexa-arm architecture may be suitable for five-input weighted spin majority gate operation, where the total spin polarization in the detector circuit is determined by the individual polarizations of all the input channels while the weight of each input signal is determined by the relative orientation of injector contact magnetization with respect to the detector.

**Conclusion**

Utilizing wafer-scale CVD graphene samples, we demonstrate a system capable of supporting the spin transport and precession with channel lengths exceeding 34 μm at room temperature. This allows us to achieve spin current communication between the magnetic memory elements in complex graphene 2D circuits of the Y-junction and hexa-arm spintronic architectures, where the spin transport is observed simultaneously in several branches of the circuit. In addition, we demonstrate that by engineering the graphene channel geometry and orientation of spin polarization, the symmetric and antisymmetric spin precession signals can be tuned in a precise manner. Our results show that commercially available CVD graphene grown on Cu foil allows for low-cost and high-quality spintronic devices on a large area, providing an attractive platform for further progress in large-scale functional spintronic circuits.

**Methods**

**Device fabrication**

The devices were prepared using CVD graphene (from Grolltex Inc) on highly doped Si (with a thermally grown 285-nm-thick $SiO_2$ layer). The graphene (Gr) channels were patterned in the Y-junction and hexa-arm structures by electron beam lithography (EBL), followed by oxygen plasma etching. The non-magnetic and magnetic contacts were patterned on Gr in two subsequent EBL steps. The non-magnetic electrodes consisted of 30 nm Au with a 10nm Ti layer for improved adhesion. Ferromagnetic contacts were prepared by electron beam evaporation of 0.6 nm of Ti, followed by *in situ* oxidation in a pure oxygen atmosphere for 10 minutes to form a $TiO_2$ tunneling barrier layer. Without exposing the device to the ambient atmosphere, in the same chamber, 40 nm of Co were deposited, after which the devices were finalized by liftoff in warm acetone at 65°C. In the final devices, the $Co/TiO_2$ contacts on Gr act as the source for spin-polarized electrons and the $n^{++}$ $Si/SiO_2$ is used as a back gate. The FM tunnel contact resistances ($R_c$), measured in a three-terminal configuration, were around 3-30 kΩ.

**Electrical measurements**

The spin transport measurements were performed in a cryostat with the sample rotation stage and variable magnetic field. The sample was kept in a vacuum and at room temperature. In the experiments, a spin injection bias current was applied using a Keithley 6221 current source, and the nonlocal voltage was detected using a Keithley 2182A nanovoltmeter; the gate voltage was applied with the $SiO_2/Si$ back gate using a Keithley 2400 multimeter.

**Data availability**

The data that support the findings of this study are available from the corresponding authors on a reasonable request.

**Acknowledgments**
The authors acknowledge financial support from the European Union's Horizon 2020 research and innovation programme under grant agreements no. 696656 and no. 785219 (Graphene Flagship Core 1 and Core 2), EU FlagEra project (from Swedish Research Council VR No. 2015-06813), Swedish Research Council VR project grants (No. 2016-03658), Graphene center and the AoA Nano program at Chalmers University of Technology.

**Author affiliations**

[1]Department of Microtechnology and Nanoscience, Chalmers University of Technology, SE-41296, Göteborg, Sweden

**Contributions**
SPD and DK conceived the idea and designed the experiments. DK fabricated and characterized the devices. BK and AMH participated in graphene preparation. DK and SPD analyzed and interpreted the experimental data, compiled the figures and wrote the manuscript. SPD supervised the research project.

**Competing interests**
The authors declare no competing financial interests.

**Corresponding authors**:
Correspondence should be addressed to
Dmitrii Khokhriakov, Email: dmikho@chalmers.se;
Saroj P. Dash, Email: saroj.dash@chalmers.se

**References**

[1]     Albert Fert, Nobel Lecture: Origin, development, and future of spintronics, Rev. Mod. Phys. (2008). https://doi.org/10.1109/JPROC.2016.2554518.

[2]     I. Žutić, J. Fabian, S. Das Sarma, Spintronics: Fundamentals and applications, Rev. Mod. Phys. 76 (2004) 323–410. https://doi.org/10.1103/RevModPhys.76.323.

[3]     H. Dery, P. Dalal, Ł. Cywiński, L.J. Sham, Spin-based logic in semiconductors for reconfigurable large-scale circuits, Nature. 447 (2007) 573. https://doi.org/10.1038/nature05833.

[4]     B. Behin-Aein, D. Datta, S. Salahuddin, S. Datta, Proposal for an all-spin logic device with built-in memory, Nat. Nanotechnol. 5 (2010) 266. https://doi.org/10.1038/nnano.2010.31.

[5]     X. Lin, L. Su, Z. Si, Y. Zhang, A. Bournel, Y. Zhang, J.-O. Klein, A. Fert, W. Zhao, Gate-Driven Pure Spin Current in Graphene, Phys. Rev. Appl. 8 (2017) 34006. https://doi.org/10.1103/PhysRevApplied.8.034006.

[6]     H. Dery, H. Wu, B. Ciftcioglu, M. Huang, Y. Song, R. Kawakami, J. Shi, I. Krivorotov, I. Zutic, L.J. Sham, Nanospintronics Based on Magnetologic Gates, IEEE Trans. Electron

Devices. 59 (2012) 259–262. https://doi.org/10.1109/TED.2011.2173498.

[7]     H. Wen, H. Dery, W. Amamou, T. Zhu, Z. Lin, J. Shi, I. Žutić, I. Krivorotov, L.J. Sham, R.K. Kawakami, Experimental demonstration of XOR operation in graphene magnetologic gates at room temperature, Phys. Rev. Appl. 5 (2016) 44003.

[8]     J. Ingla-Aynés, A.A. Kaverzin, B.J. van Wees, Carrier Drift Control of Spin Currents in Graphene-Based Spin-Current Demultiplexers, Phys. Rev. Appl. 10 (2018) 44073. https://doi.org/10.1103/PhysRevApplied.10.044073.

[9]     Y. Fukuma, L. Wang, H. Idzuchi, S. Takahashi, S. Maekawa, Y. Otani, Giant enhancement of spin accumulation and long-distance spin precession in metallic lateral spin valves, Nat. Mater. 10 (2011) 527.

[10]    X. Lou, C. Adelmann, S.A. Crooker, E.S. Garlid, J. Zhang, K.S.M. Reddy, S.D. Flexner, C.J. Palmstrøm, P.A. Crowell, Electrical detection of spin transport in lateral ferromagnet–semiconductor devices, Nat. Phys. 3 (2007) 197.

[11]    S.P. Dash, S. Sharma, R.S. Patel, M.P. de Jong, R. Jansen, Electrical creation of spin polarization in silicon at room temperature, Nature. 462 (2009) 491.

[12]    M. Oltscher, F. Eberle, T. Kuczmik, A. Bayer, D. Schuh, D. Bougeard, M. Ciorga, D. Weiss, Gate-tunable large magnetoresistance in an all-semiconductor spin valve device, Nat. Commun. 8 (2017) 1807. https://doi.org/10.1038/s41467-017-01933-2.

[13]    A. Spiesser, H. Saito, Y. Fujita, S. Yamada, K. Hamaya, S. Yuasa, R. Jansen, Giant Spin Accumulation in Silicon Nonlocal Spin-Transport Devices, Phys. Rev. Appl. 8 (2017) 64023. https://doi.org/10.1103/PhysRevApplied.8.064023.

[14]    A. Dankert, J. Geurs, M.V. Kamalakar, S. Charpentier, S.P. Dash, Room Temperature Electrical Detection of Spin Polarized Currents in Topological Insulators, Nano Lett. 15 (2015) 7976–7981. https://doi.org/10.1021/acs.nanolett.5b03080.

[15]    W. Han, R.K. Kawakami, M. Gmitra, J. Fabian, Graphene spintronics, Nat. Nanotechnol. 9 (2014) 794. https://doi.org/10.1038/nnano.2014.214.

[16]    S. Roche, J. Åkerman, B. Beschoten, J.-C. Charlier, M. Chshiev, S. Prasad Dash, B.

Dlubak, J. Fabian, A. Fert, M. Guimarães, F. Guinea, I. Grigorieva, C. Schönenberger, P. Seneor, C. Stampfer, S.O. Valenzuela, X. Waintal, B. van Wees, Graphene spintronics: the European Flagship perspective, 2D Mater. 2 (2015) 030202. https://doi.org/10.1088/2053-1583/2/3/030202.

[17] N. Tombros, C. Jozsa, M. Popinciuc, H.T. Jonkman, B.J. van Wees, Electronic spin transport and spin precession in single graphene layers at room temperature, Nature. 448 (2007) 571–574. https://doi.org/10.1038/nature06037.

[18] B. Dlubak, M.-B. Martin, C. Deranlot, B. Servet, S. Xavier, R. Mattana, M. Sprinkle, C. Berger, W.A. De Heer, F. Petroff, A. Anane, P. Seneor, A. Fert, Highly efficient spin transport in epitaxial graphene on SiC, Nat. Phys. 8 (2012) 557. https://doi.org/10.1038/nphys2331.

[19] B. Raes, J.E. Scheerder, M. V Costache, F. Bonell, J.F. Sierra, J. Cuppens, J. Van de Vondel, S.O. Valenzuela, Determination of the spin-lifetime anisotropy in graphene using oblique spin precession, Nat. Commun. 7 (2016) 11444. https://doi.org/10.1038/ncomms11444.

[20] M. Drögeler, C. Franzen, F. Volmer, T. Pohlmann, L. Banszerus, M. Wolter, K. Watanabe, T. Taniguchi, C. Stampfer, B. Beschoten, Spin Lifetimes Exceeding 12 ns in Graphene Nonlocal Spin Valve Devices, Nano Lett. (2016). https://doi.org/10.1021/acs.nanolett.6b00497.

[21] J. Ingla-Aynés, M.H.D. Guimarães, R.J. Meijerink, P.J. Zomer, B.J. van Wees, 24-um spin relaxation length in boron nitride encapsulated bilayer graphene, Phys. Rev. B. 92 (2015) 201410. https://doi.org/10.1103/PhysRevB.92.201410.

[22] A. Avsar, T.-Y. Yang, S. Bae, J. Balakrishnan, F. Volmer, M. Jaiswal, Z. Yi, S.R. Ali, G. Güntherodt, B.H. Hong, B. Beschoten, B. Özyilmaz, Toward Wafer Scale Fabrication of Graphene Based Spin Valve Devices, Nano Lett. 11 (2011) 2363–2368. https://doi.org/10.1021/nl200714q.

[23] I.G. Serrano, J. Panda, F. Denoel, Ö. Vallin, D. Phuyal, O. Karis, M.V. Kamalakar, Two-Dimensional Flexible High Diffusive Spin Circuits, Nano Lett. 19 (2019) 666–673.

https://doi.org/10.1021/acs.nanolett.8b03520.

[24] M.V. Kamalakar, C. Groenveld, A. Dankert, S.P. Dash, Long distance spin communication in chemical vapour deposited graphene, Nat. Commun. 6 (2015) 6766. https://doi.org/10.1038/ncomms7766.

[25] B. Zhao, D. Khokhriakov, B. Karpiak, A.M. Hoque, L. Xu, L. Shen, Y.P. Feng, X. Xu, Y. Jiang, S.P. Dash, Electrically controlled spin-switch and evolution of Hanle spin precession in graphene, 2D Mater. 6 (2019) 35042. https://doi.org/10.1088/2053-1583/ab1d83.

[26] W. Amamou, G. Stecklein, S.J. Koester, P.A. Crowell, R.K. Kawakami, Spin Absorption by In Situ Deposited Nanoscale Magnets on Graphene Spin Valves, Phys. Rev. Appl. 10 (2018) 44050. https://doi.org/10.1103/PhysRevApplied.10.044050.

[27] F.J. Jedema, H.B. Heersche, A.T. Filip, J.J.A. Baselmans, B.J. van Wees, Electrical detection of spin precession in a metallic mesoscopic spin valve, Nature. 416 (2002) 713–716. https://doi.org/10.1038/416713a.

[28] Z.M. Gebeyehu, S. Parui, J.F. Sierra, M. Timmermans, M.J. Esplandiu, S. Brems, C. Huyghebaert, K. Garello, M. V Costache, S.O. Valenzuela, Spin communication over 30 µm long channels of chemical vapor deposited graphene on SiO2, 2D Mater. 6 (2019) 34003. https://doi.org/10.1088/2053-1583/ab1874.

[29] M. Drögeler, F. Volmer, M. Wolter, B. Terrés, K. Watanabe, T. Taniguchi, G. Güntherodt, C. Stampfer, B. Beschoten, Nanosecond Spin Lifetimes in Single- and Few-Layer Graphene–hBN Heterostructures at Room Temperature, Nano Lett. 14 (2014) 6050–6055. https://doi.org/10.1021/nl501278c.

[30] J. Xu, S. Singh, J. Katoch, G. Wu, T. Zhu, I. Žutić, R.K. Kawakami, Spin inversion in graphene spin valves by gate-tunable magnetic proximity effect at one-dimensional contacts, Nat. Commun. 9 (2018) 2869. https://doi.org/10.1038/s41467-018-05358-3.

[31] B. Karpiak, A. Dankert, A.W. Cummings, S.R. Power, S. Roche, S.P. Dash, 1D ferromagnetic edge contacts to 2D graphene/h-BN heterostructures, 2D Mater. 5 (2017) 14001. https://doi.org/10.1088/2053-1583/aa8d2b.

[32] M.V. Kamalakar, A. Dankert, J. Bergsten, T. Ive, S.P. Dash, Enhanced Tunnel Spin Injection into Graphene using Chemical Vapor Deposited Hexagonal Boron Nitride, Sci. Rep. 4 (2014) 6146. https://doi.org/10.1038/srep06146.

[33] A. Dankert, P. Pashaei, M.V. Kamalakar, A.P.S. Gaur, S. Sahoo, I. Rungger, A. Narayan, K. Dolui, M.A. Hoque, R.S. Patel, M.P. de Jong, R.S. Katiyar, S. Sanvito, S.P. Dash, Spin-Polarized Tunneling through Chemical Vapor Deposited Multilayer Molybdenum Disulfide, ACS Nano. 11 (2017) 6389–6395. https://doi.org/10.1021/acsnano.7b02819.

[34] A. Dankert, M. Venkata Kamalakar, A. Wajid, R.S. Patel, S.P. Dash, Tunnel magnetoresistance with atomically thin two-dimensional hexagonal boron nitride barriers, Nano Res. 8 (2015) 1357–1364. https://doi.org/10.1007/s12274-014-0627-4.

[35] M.V. Kamalakar, A. Dankert, J. Bergsten, T. Ive, S.P. Dash, Spintronics with graphene-hexagonal boron nitride van der Waals heterostructures, Appl. Phys. Lett. 105 (2014) 212405. https://doi.org/10.1063/1.4902814.

[36] M.V. Kamalakar, A. Dankert, P.J. Kelly, S.P. Dash, Inversion of Spin Signal and Spin Filtering in Ferromagnet|Hexagonal Boron Nitride-Graphene van der Waals Heterostructures, Sci. Rep. 6 (2016) 21168. https://doi.org/10.1038/srep21168.

[37] A. Dankert, S.P. Dash, Electrical gate control of spin current in van der Waals heterostructures at room temperature, Nat. Commun. 8 (2017) 16093. https://doi.org/10.1038/ncomms16093.

[38] W. Yan, O. Txoperena, R. Llopis, H. Dery, L.E. Hueso, F. Casanova, A two-dimensional spin field-effect switch, Nat. Commun. 7 (2016). https://doi.org/10.1038/ncomms13372.

[39] B. Karpiak, A.W. Cummings, K. Zollner, M. Vila, D. Khokhriakov, A.M. Hoque, A. Dankert, P. Svedlindh, J. Fabian, S. Roche, S.P. Dash, Magnetic proximity in a van der Waals heterostructure of magnetic insulator and graphene, 2D Mater. 7 (2019) 15026. https://doi.org/10.1088/2053-1583/ab5915.

[40] D. Khokhriakov, A.W. Cummings, K. Song, M. Vila, B. Karpiak, A. Dankert, S. Roche, S.P. Dash, Tailoring emergent spin phenomena in Dirac material heterostructures, Sci. Adv. (2018). https://doi.org/10.1126/sciadv.aat9349.

[41] C.K. Safeer, J. Ingla-Aynés, F. Herling, J.H. Garcia, M. Vila, N. Ontoso, M.R. Calvo, S. Roche, L.E. Hueso, F. Casanova, Room-Temperature Spin Hall Effect in Graphene/MoS2 van der Waals Heterostructures, Nano Lett. 19 (2019) 1074–1082. https://doi.org/10.1021/acs.nanolett.8b04368.

[42] T.S. Ghiasi, A.A. Kaverzin, P.J. Blah, B.J. van Wees, Charge-to-Spin Conversion by the Rashba–Edelstein Effect in Two-Dimensional van der Waals Heterostructures up to Room Temperature, Nano Lett. 19 (2019) 5959–5966. https://doi.org/10.1021/acs.nanolett.9b01611.

[43] A.M. Hoque, D. Khokhriakov, B. Karpiak, S.P. Dash, All-electrical creation and control of giant spin-galvanic effect in 1T-MoTe2/graphene heterostructures at room temperature, ArXiv Prepr. ArXiv1908.09367. (2019).

[44] B. Zhao, D. Khokhriakov, Y. Zhang, H. Fu, B. Karpiak, A.M. Hoque, X. Xu, Y. Jiang, B. Yan, S.P. Dash, Observation of Spin Hall Effect in Semimetal WTe2, ArXiv Prepr. ArXiv1812.02113. (2018).

[45] B. Zhao, B. Karpiak, D. Khokhriakov, A.M. Hoque, X. Xu, Y. Jiang, S.P. Dash, Edelstein Effect in Type-II Weyl Semimetal WTe2 up to Room Temperature, ArXiv Prepr. ArXiv1910.06225. (2019).

[46] D. Khokhriakov, A.M. Hoque, B. Karpiak, S.P. Dash, Giant and Gate-tunable Spin-Galvanic Effect in Graphene Topological insulator van der Waals Heterostructures at Room Temperature, ArXiv Prepr. ArXiv1910.06760. (2019).

[47] J.S. Lee, S.H. Choi, S.J. Yun, Y.I. Kim, S. Boandoh, J.-H. Park, B.G. Shin, H. Ko, S.H. Lee, Y.-M. Kim, Y.H. Lee, K.K. Kim, S.M. Kim, Wafer-scale single-crystal hexagonal boron nitride film via self-collimated grain formation, Science (80-. ). 362 (2018) 817 LP – 821. https://doi.org/10.1126/science.aau2132.

[48] H. Yu, M. Liao, W. Zhao, G. Liu, X.J. Zhou, Z. Wei, X. Xu, K. Liu, Z. Hu, K. Deng, S. Zhou, J.-A. Shi, L. Gu, C. Shen, T. Zhang, L. Du, L. Xie, J. Zhu, W. Chen, R. Yang, D. Shi, G. Zhang, Wafer-Scale Growth and Transfer of Highly-Oriented Monolayer MoS2 Continuous Films, ACS Nano. 11 (2017) 12001–12007.

https://doi.org/10.1021/acsnano.7b03819.

[49]  T. Kimura, Y. Otani, P.M. Levy, Electrical Control of the Direction of Spin Accumulation, Phys. Rev. Lett. 99 (2007) 166601. https://doi.org/10.1103/PhysRevLett.99.166601.

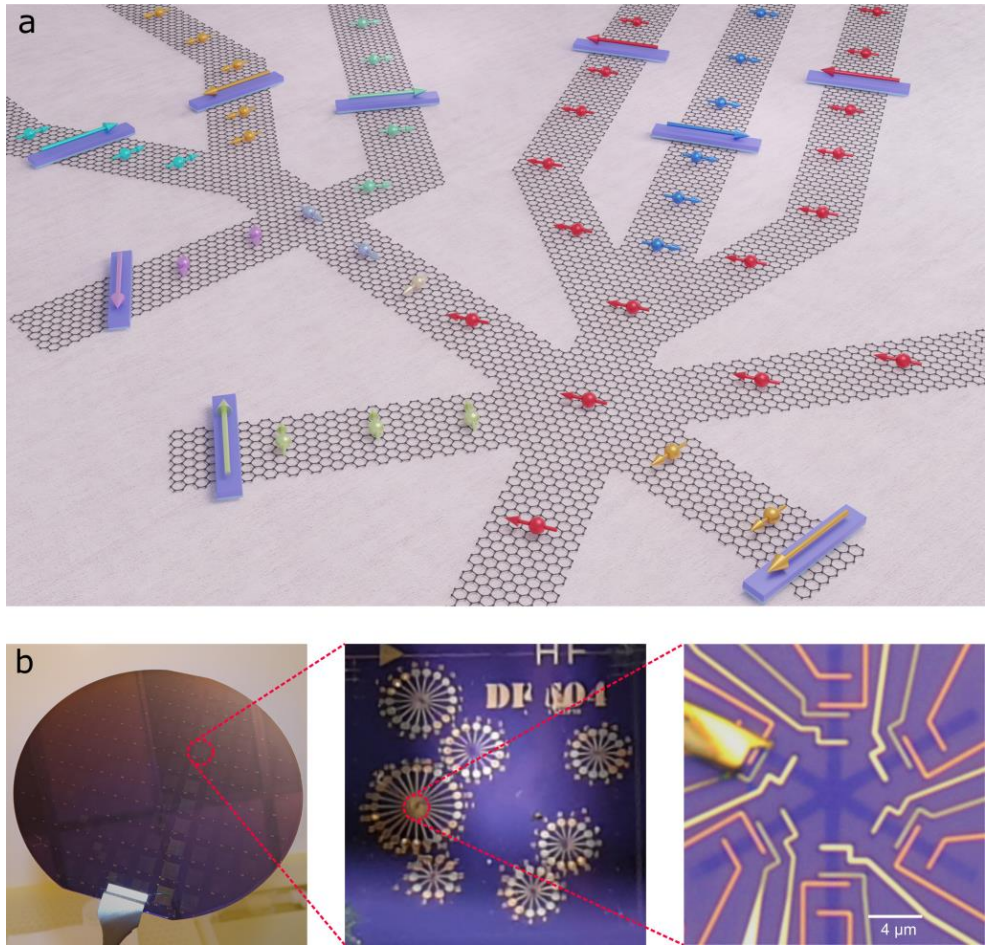

**Figure 1. The 2D graphene spintronic circuit. a.** An artistic representation of a graphene spin circuit composed of multiple nanomagnetic elements interconnected with graphene channels. The functionality of the circuit includes spin injection, transport, detection in graphene and magnetization dynamics of the nanomagnets. **b.** The nanofabricated spin circuit (a hexa-arm architecture) on a chip (7x7 mm$^2$) picked from a 4-inch CVD graphene wafer on SiO$_2$/Si substrate. The circuit shows the patterned CVD graphene channels with ferromagnetic TiO$_2$/Co (yellow) tunnel contacts and non-magnetic Ti/Au (orange) reference contacts defined by electron beam lithography. The ferromagnetic elements are used for injection and detection of spin-polarized current in the graphene circuit.

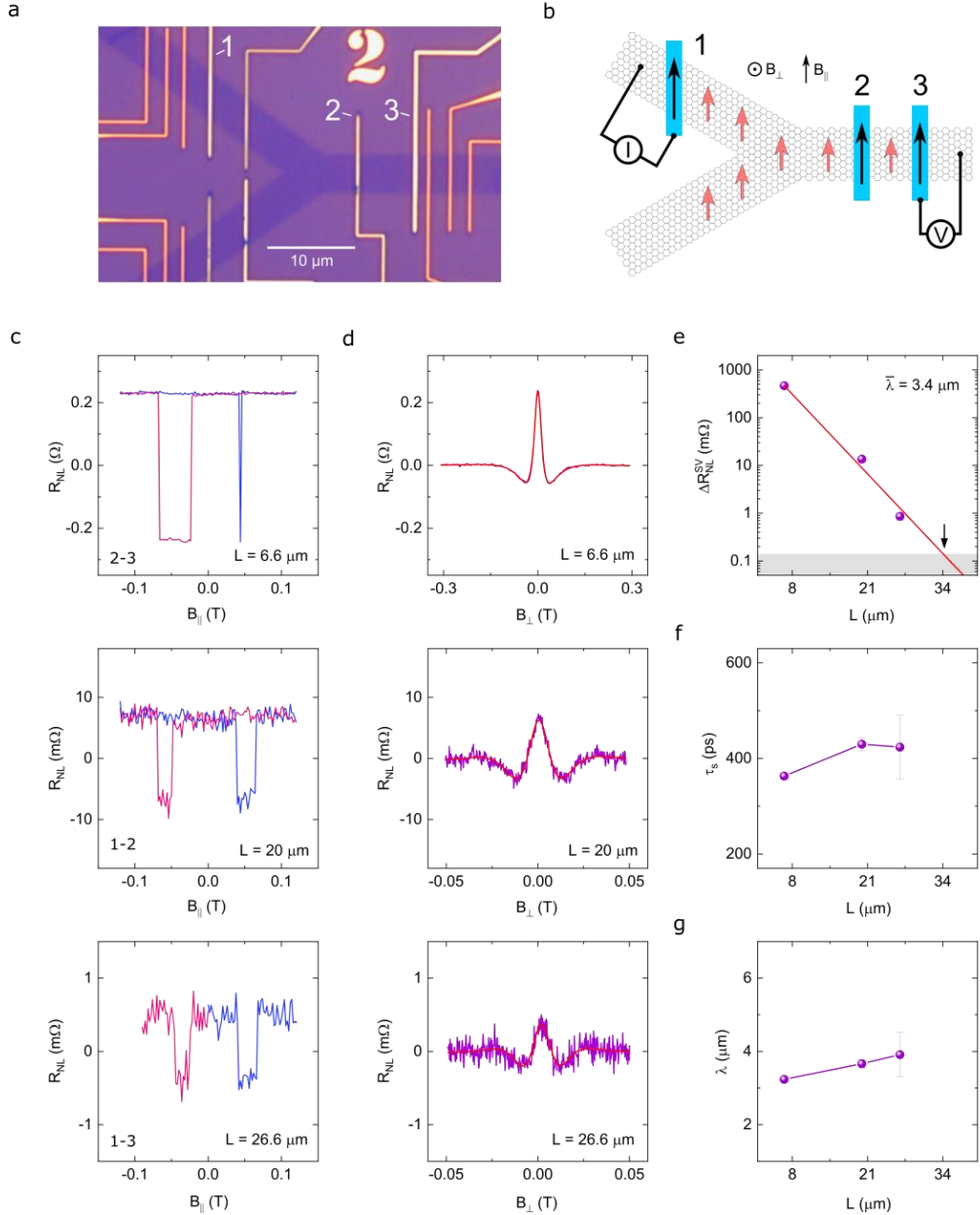

**Figure 2**. **Graphene Y-junction spin circuit**. **a-b.** Optical microscope image and schematic of a fabricated CVD graphene device in a Y-junction spin circuit having spin transport channel lengths $L$ = 6.6, 20 and 26.6 μm with multiple ferromagnetic Co/TiO$_2$ (yellow) injector/detector and non-magnetic Au/Ti (orange) reference contacts. The NL measurement scheme is shown with current (*I*) and voltage (*V*) circuits. The first geometry (between contacts 2 and 3) is a conventional longitudinal channel, whereas the other two channels are in a Y-junction circuit. **c.** NL spin valve signal for different graphene channel lengths with in-plane magnetic field ($B_∥$) sweep. **d.** Hanle
19

spin precession signals $R_{NL}$ measured with a perpendicular magnetic field ($B_\perp$) sweep. The data shown for $L$ = 20 and 26.6 μm are obtained by averaging the Hanle signals in parallel (↑↑) and antiparallel (↑↓) configuration of ferromagnetic electrodes as $R_{NL} = \frac{R_{NL}^{\uparrow\uparrow} - R_{NL}^{\uparrow\downarrow}}{2}$. Measurements were performed with bias currents $I$ = 50 - 300 μA. The data points are fitted with the Hanle equation to extract spin lifetime $\tau_s$ and spin diffusion constant $D_s$. **e.** Channel length dependence of the spin valve magnitude. The arrow shows the estimated maximum spin transport distance found at the intersection of the fit to an exponential decrease of spin valve amplitude with distance (red line) and the root mean square noise level of the measurements (shaded area). **f-g.** Channel length dependence of spin lifetime $\tau_s$ and spin diffusion length $\lambda_s$ extracted from Hanle spin precession measurements.

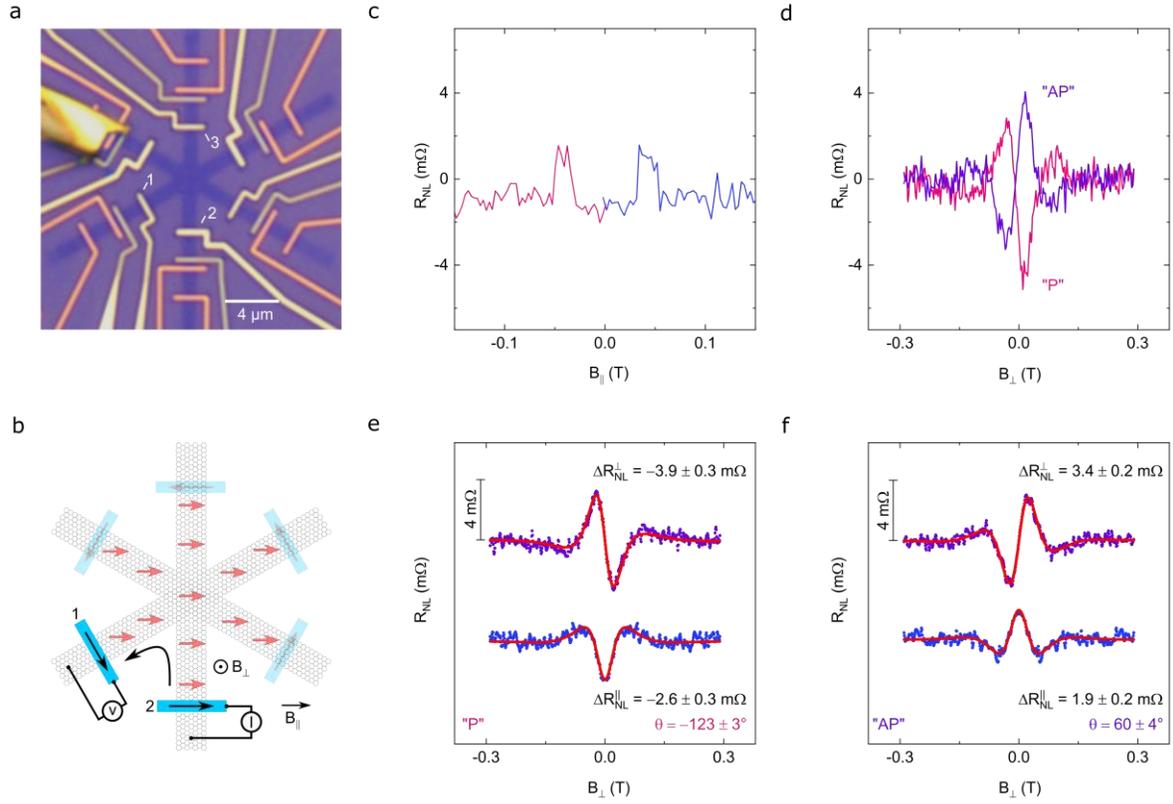

**Figure 3**. **Graphene hexa-arm spin circuit architecture**. **a.** Optical microscopy image of a fabricated CVD graphene device in a hexa-arm junction having multiple ferromagnetic tunnel contacts of Co/TiO$_2$ with the channel length between inner electrodes of $L = 7.8$ μm. **b.** A schematic of the hexa-arm channel with a 60° angle between graphene branches and the injector/detector contacts placed perpendicularly to the respective graphene channels. **c.** NL spin valve signal for the measurement configuration highlighted in (b) with in-plane magnetic field ($B_∥$) sweep. **d.** Asymmetric NL Hanle spin precession signal obtained by a perpendicular magnetic field ($B_⊥$) sweep in "parallel" and "antiparallel" configurations of ferromagnetic electrodes at room temperature. The measurements were performed with the bias current $I = 300$ μA and back gate voltage $V_g = -80$ V. **e-f.** Hanle signals deconvoluted into the antisymmetric and symmetric components with fits to eq. 1. The curves are offset vertically for clarity.

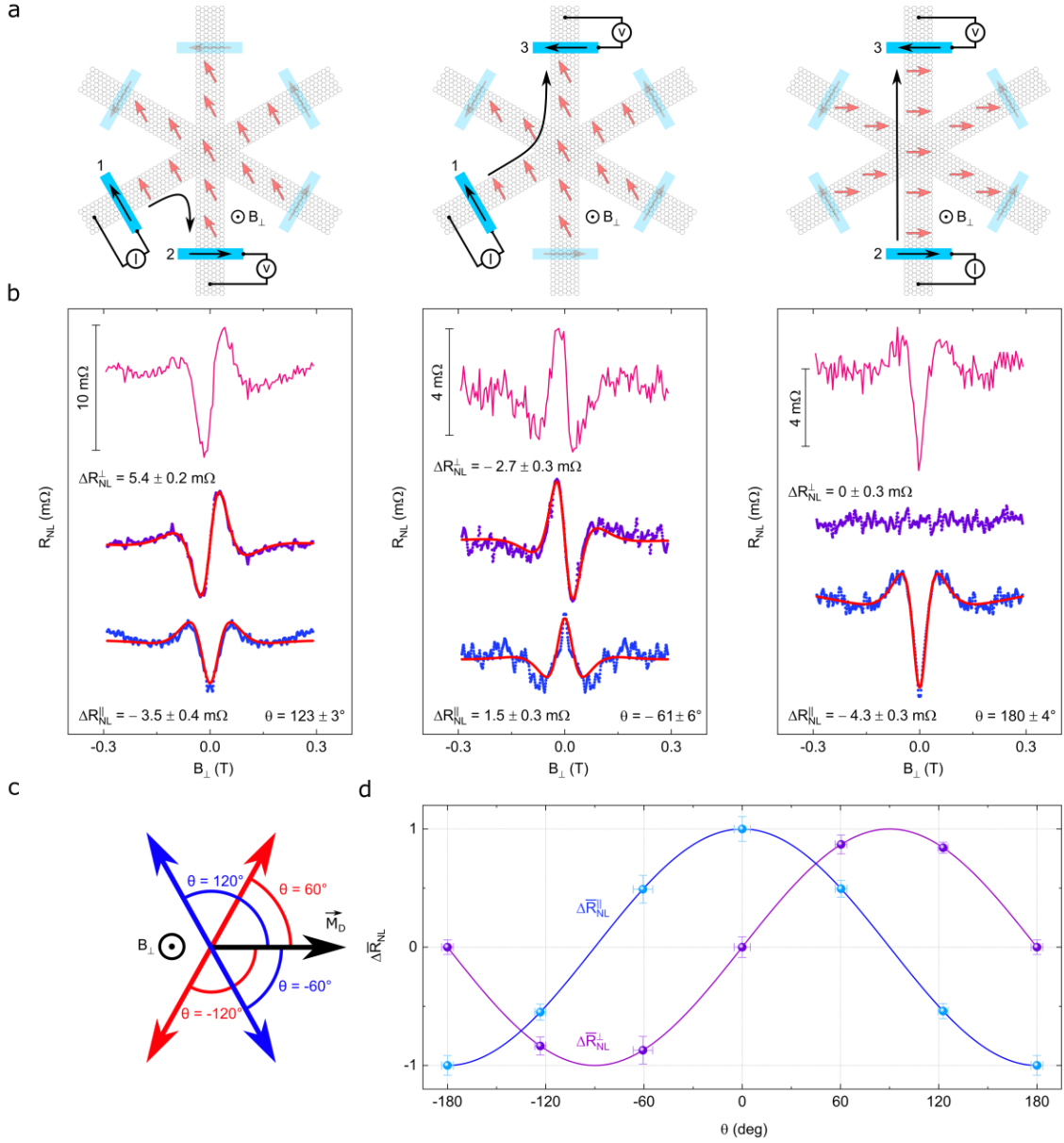

**Figure 4. Non-collinear spin precession in the hexa-arm device. a.** Schematics of the hexa-arm graphene spin circuit architecture with the detector ferromagnetic contacts placed at different angles with respect to the injector ferromagnet (60°, 120°, 0°) in different graphene branches. **b.** Hanle spin precession curves (pink) and their deconvoluted antisymmetric (purple) and symmetric (blue) components measured in the outlined channels. The measurements were performed with the bias current $I = 300$ μA and the back gate voltage $V_g = -80$ V. The curves are offset vertically for clarity. **c.** Schematic of the angle notation used in all graphs with the black arrow representing the detector magnetization direction $\vec{M}_D$ and the red and blue arrows showing the injector

magnetization. **d.** The simulated amplitude of symmetric and antisymmetric Hanle components as a function of the angle between injector/detector contacts. The points show the normalized experimental data.